# Berry curvature and symmetry broken induced Hall effect in $MoS_2$

Michael Zhang[1] and Gerry West

School of Physics and Astronomy, University of Minnesota,

Minneapolis, MN, 55455

Inversion symmetry breaking and spin-orbit coupling lead to valley and spin Hall effect in $MoS_2$. Because of the large valley separation in momentum space, the valley index is expected to be robust. In this paper, quantum Hall effect in $MoS_2$ originated from Berry curvature is analyzed after review of symmetry structure and spin-orbit coupling Hamiltonian of $MoS_2$. Finally an expression and rough calculation is given for valley and spin Hall effect.

## I. Barry curvature in Bloch bands

If the Hamiltonian $H$ is independent of time, then a particle which starts out in the $n^{th}$ eigenstate $\psi_n$

$$H\psi_n = E_n\psi_n \qquad (1)$$

Remains in the $n^{th}$ eigenstate, simply picking up a phase factor

$$\Psi_n(t) = \psi_n e^{-\frac{iE_n}{\hbar}t} \qquad (2)$$

If the Hamiltonian changes with time $H(t)$, according to the Adiabatic Theorem, $\Psi_n(t)$ remains in the $n^{th}$ eigenstate of the evolving Hamiltonian with several more phase factors.

---

[1] Corresponding author email: zhanx015@umn.edu

$$\Psi_n(t) = e^{i\gamma_n(t)} \exp\left[-\frac{i}{\hbar}\int_0^t dt' E_n(t')\right]\psi_n \tag{3}$$

$\exp\left[-\frac{i}{\hbar}\int_0^t dt' E_n(t')\right]$ is known as dynamic phase and $\gamma_n$ is known as geometric phase (since it can be expressed as a geometric path integral) or Berry phase. It can be proven for a closed path, $\gamma_n$ become a gauge-invariant physical quantity.

$$\gamma_n = i\oint_C d\vec{R}\,\langle\psi_n|\nabla_R\psi_n\rangle \tag{4}$$

With $\vec{R}(t)$ is the parameter in Hamiltonian changing with time (Details see Ref. [1]). In analogy to electrodynamics, we can define a "gauge field" $\Omega_n$, called Berry curvature and Berry phase can therefore be written as a surface integral

$$\Omega_n = i\nabla_R\times\langle\psi_n|\nabla_R\psi_n\rangle \tag{5a}$$

$$\gamma_n = \int_S d\vec{s}\,\Omega_n \tag{5b}$$

From our class, we know band structure in a periodic potential $V(\vec{r}) = V(\vec{r}+\vec{a})$ is given by Bloch's theorem

$$\psi_{n\vec{k}}(\vec{r}+\vec{a}) = e^{i\vec{k}\cdot\vec{a}}\psi_{n\vec{k}}(\vec{r}) \tag{6a}$$

$$\psi_{n\vec{k}}(\vec{r}) = e^{i\vec{k}\cdot\vec{r}}u_{n\vec{k}}(\vec{r}) \tag{6b}$$

$\vec{a}$ is the Bravais lattice vector and $u_{n\vec{k}}$ is the Bloch periodic function satisfy $u_{n\vec{k}}(\vec{r}) = u_{n\vec{k}}(\vec{r}+\vec{a})$. Since $\vec{k}$ dependence of the basis function is inherent to the Bloch problem, various Berry phase effects are expected in crystals. [2] For example, if $\vec{k}$ is forced to vary in the momentum space, then the Bloch state will have a Berry phase if the integral path C in equation(4) is closed.

There are two ways to generate a close path in momentum space. One can apply a magnetic field introducing a cyclotron motion along a closed orbit in the $\vec{k}$ space, such effect has been observed in 2D graphene system[3].

The Berry curvature $\Omega_n(\vec{k})$ is an intrinsic property of the band structure because it only depends on the wave function. It is nonzero in a wide range of materials, in particular, crystals with broken time-reversal or inversion symmetry. We will encounter this again in discussing anomalous Hall effect of $MoS_2$. Another way is to apply an electric field $\vec{\varepsilon}$, a closed loop will be realized in within the Brillouin zone. In this case, the Berry phase across the Brillouin zone is called Zak's phase [2]

$$\gamma_n = \oint_{BZ} d\vec{k} \langle u_n(\vec{k}) | i\nabla_{\vec{k}} | u_n(\vec{k}) \rangle \tag{7}$$

This phase depends on the topology of Brillouin zone. In the presence of the electric field, electron can acquire an anomalous velocity proportional to the Berry curvature of the band. This velocity is responsible for transport such as various Hall effect. The average velocity $\vec{v}_n(\vec{k})$ in a state $\vec{k}$ is found [2]

$$\vec{v}_n(\vec{k}) = \frac{\partial E_n(\vec{k})}{\hbar \partial \vec{k}} - \frac{e}{\hbar} \vec{\varepsilon} \times \Omega_n(\vec{k}) \tag{8}$$

Equation(8) reveals that, in addition to the band energy, the Berry curvature of Bloch bands is also required for a complete description of electron dynamics. However, conventional first term

$\frac{\partial E_n(\vec{k})}{\hbar \partial \vec{k}}$ has been successful in describing electron velocity, it is important to consider when the second term with Berry curvature $-\frac{e}{\hbar}\vec{\varepsilon} \times \Omega_n(\vec{k})$ cannot be neglected.

The velocity should be invariant under time-reversal and spatial inversion operations. Under time reversal, $\vec{v}_n$ and $\vec{k}$ change the sign while $\vec{\varepsilon}$ stay fixed. Under spatial inversion, $\vec{v}_n$, $\vec{k}$ and $\vec{\varepsilon}$ all change the sign. Therefore if the system has time-reversal symmetry, $\Omega_n(-\vec{k}) = -\Omega_n(\vec{k})$ should be satisfied in equation(8), whereas for spatial inversion system $\Omega_n(-\vec{k}) = \Omega_n(\vec{k})$ is required.

As a result, for crystals with simultaneous time-reversal and spatial inversion symmetry, the Berry curvature vanishes identically throughout the Brillouin zone. However, for systems with either time-reversal or inversion symmetries broken, Berry curvature term might be important. Graphene single layer with an external electrical field is one of such systems. As an example, the external electrical filed can induce a Rashba term which open a band gap and breaks the inversion symmetry.

## II. $MoS_2$ electronic structure and Hamiltonian

However, the intrinsic graphene does not have a nonzero band gap and limit the use in electronic devices. This prompted research in materials similar to graphene, such as, 2D monolayer $MoS_2$.

In monolayer $MoS_2$, the conduction and valence band edges are located at the corners (K points) of the 2D hexagonal plane. [4]Similar to graphene, the two inequivalent valleys constitue a binary index for low energy carriers. Because of the large valley separation in momentum space, the valley index is expected to be robust against scattering by smooth deformations and long wavelength phonons[5]. Hence we may investigate coexistence of spin and valley Hall effect.

$MoS_2$ monolayers have two important distinctions from graphene. First, inversion symmetry is explicitly broken in monolayer $MoS_2$, which can give rise to the valley Hall effect where carriers in different valleys flow to opposite transverse edges when an in-plane electric field is applied. Second, $MoS_2$ has a strong spin-orbit coupling originated from the d orbitals of the heavy metal atoms, and can be an interesting platform to explore spin physics and spintronics applications absent in graphene due to its weak intrinsic Dresselhaus spin-orbit coupling[6].

In its bulk form, $MoS_2$ has the stacking order with the space group $D_{6h}^4$, which is inversion symmetric. When it is thinned down to a monolayer, the crystal symmetry reduces to $D_{3h}^1$, and inversion symmetry is explicitly broken: taking the Mo atom as the inversion center, an S atom will be mapped onto an empty location. As a consequence, the effects are expected only in thin films with odd number of layers.

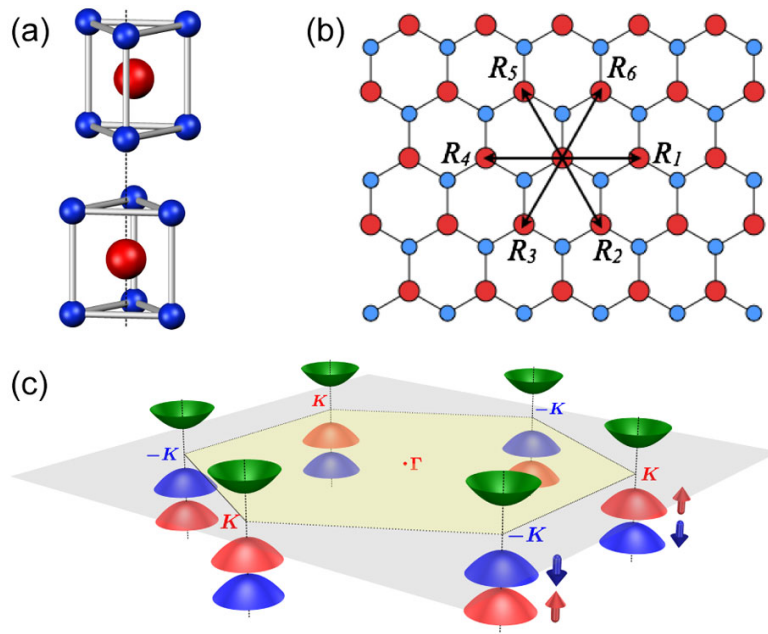

FIG.1. (a) The unit cell of bulk 2H-MoS2, which has the inversion center located in the middle plane. It contains two unit cells of MoS2 monolayers, which lacks an inversion center. (b) Top view of the MoS2 monolayer. Ri are the vectors connecting nearest Mo atoms. (c) Schematic drawing of the band structure at the band edges located at the K points.[5]

Minimal band model on the basis of general symmetry can be constructed. The band structure of $MoS_2$, to a first approximation, consists of partially filled Mo d bands lying between Mo-S sp bonding and antibonding bands. The trigonal prismatic coordination of the Mo atom splits its d orbitals into three groups: $A_1(d_{z^2})$, $E(d_{xy}, d_{x^2-y^2})$ and $E'(d_{xz}, d_{yz})$. In the monolayer limit, the reflection symmetry in the z direction permits hybridization only between $A_1$ and $E$ orbitals, which opens a band gap at the K and –K points.[7] The group of the wave vector at the band edges K is $C_{3h}$ and the symmetry adapted basis functions are

$$|\phi_c\rangle = |d_{z^2}\rangle \qquad (9a)$$

$$|\phi_v^\pm\rangle = \frac{1}{\sqrt{2}}(|d_{x^2-y^2}\rangle \pm i|d_{xy}\rangle) \qquad (9b)$$

$\pm$ is the valley index notation, $\phi_v^+$ and $\phi_v^-$ are related by time-reversal operation. Similar to (massive) graphene, to the first order in k, the $C_{3h}$ symmetry has the two-band k.p Hamiltonian

$$H_0 = bt(k_x\sigma_x + k_y\sigma_y) + \frac{E_g}{2}\sigma_z \qquad (10)$$

$\vec{\sigma}$ denotes the Pauli matrices for the two basis functions, b is the lattice constant, $t$ is the nearest neighbor effective hopping integral, and $E_g$ is the energy gap. These parameters can be found from literature by fitting to first-principle band structure calculation and listed in Table I[8] and Ref [9-10]. What distinguishes $MoS_2$ from graphene is the strong spin-orbit coupling originated from the metal d orbitals.

The conduction band-edge state is made of $d_{z^2}$ orbitals and remains spin degenerate at K points, whereas the valence-band-edge state splits. Approximating the spin-orbit coupling by the intra-atomic contribution $\vec{L} \cdot \vec{S}$, the total Hamiltonian given by

$$H = bt(k_x\sigma_x + k_y\sigma_y) + \frac{E_g}{2}\sigma_z - \Delta_{so}\frac{\sigma_z - 1}{2}s_z \qquad (11)$$

$2\Delta_{so}$ is the spin splitting at the valence band top caused by the spin-orbit coupling and $s_z$ is the Pauli matrix for true spin. The spin splitting is a general consequence of inversion symmetry breaking and does not depend on model details. This is similar to Dresselhaus spin splitting in zincblende semiconductors. [5] The eigenvalue in equation is of the form $E = \pm\sqrt{\left(\hbar v_f|\vec{k}|\right)^2 + \Delta_{so}^2}$, which also describes graphene with Rashba spin-orbit coupling.

|        | $b(A)$ | $E_g(eV)$ | $t(eV)$ | $2\Delta_{so}(eV)$ |
|--------|--------|-----------|---------|--------------------|
| $MoS_2$  | 3.193  | 1.66      | 1.10    | 0.15               |
| $WS_2$   | 3.197  | 1.79      | 1.37    | 0.43               |
| $MoSe_2$ | 3.313  | 1.47      | 0.94    | 0.18               |
| $WSe_2$  | 3.310  | 1.60      | 1.19    | 0.46               |

TABLE.I Fitting result from first-principles band structure calculation[8]

### III. Valley/ Spin Hall effects in $MoS_2$

The quantum Hall effect find 2D electron gas can be exactly quantized in unit of $\frac{e^2}{\hbar}$. This effect has been observed in graphene under room temperature[3] and blossomed into an important research topic in condensed-matter physics. For systems with either time-reversal or inversion symmetries broken such as $MoS_2$, as discussed in part I and II, Berry curvature become pronounced and able to drive valley Hall and spin Hall.

In the presence of an in-plane electric field, an electron will acquire an anomalous velocity proportional to the Berry curvature in the transverse direction, giving rise to an intrinsic contribution to the quantum Hall conductivity[2]

$$\sigma = \frac{e^2}{\hbar} \int \frac{d\vec{k}}{2\pi} f(\vec{k}) \Omega(\vec{k}) \tag{12}$$

Where $f(\vec{k})$ is the Fermi-Dirac distribution function and $\Omega(\vec{k})$ is the Berry curvature $\Omega_n(\vec{k}) = \hat{z} \cdot \nabla_{\vec{k}} \times \langle u_n(\vec{k}) | i \nabla_{\vec{k}} | u_n(\vec{k}) \rangle$, which is in $\hat{z}$ direction. Only consider electrons, further from Ref. [2], the $\Omega(\vec{k})$ for Dirac fermions in conduction band has the analytic expression

$$\Omega_c(\vec{k}) = \mp \frac{2b^2 t^2 E_g'}{\left(E_g'^2 + 4b^2 t^2 k^2\right)^{\frac{3}{2}}} \tag{13a}$$

In valence band,

$$\Omega_v(\vec{k}) = -\Omega_c(\vec{k}) \tag{13b}$$

Berry curvatures have opposite sign in opposite valleys. In the same valley, the Berry curvature depend on effective band gap $E_g' = E_g \mp s_z \Delta_{so}$. Near the vicinity of K point, $E_g \gg btk$, the curvature $\Omega_{c,K}$ is nearly constant.

Now try to roughly evaluate valley Hall conductivity for electrons $\sigma_v^e$ with equation(12), i.e. $MoS_2$ is n-type doped . The derivation is assumed zero Kelvin near K valley

$$\sigma_v^e = \frac{e^2}{\hbar} \int_{\frac{E_g}{2}}^{\infty} dE\, g(E) f(E)\, \Omega_{c,K} \approx \frac{e^2}{\hbar} \int_{\frac{E_g}{2}}^{E_f} dE\, \frac{2}{\pi(\hbar v_f)^2} E \frac{2b^2 t^2}{E_g'^2}$$

$$= \frac{2b^2 t^2}{\pi \hbar^2 v_f^2 E_g'^2} \left( E_f^2 - \frac{E_g^2}{4} \right) \frac{e^2}{\hbar} \tag{14}$$

$g(E) = \frac{2}{\pi(\hbar v_f)^2} E$ is the electron density of states at K point, $f(E)$ become delta function at zero Kelvin, $E_f$ is the Fermi energy at conduction band minimum. If the $MoS_2$ is doped so that $E_f = 2eV$, $v_f = 10^7 cm/s$, with other parameters in TABLE. I, $\sigma_v^e$ is calculated to be (in unit of $\frac{e^2}{\hbar}$)

$$\sigma_v^e = \frac{2(3.193 \times 10^{-8})^2 (1.1)^2 \left(2^2 - \frac{1.66^2}{4}\right)}{\pi (6.1581 \times 10^{-16})^2 (10^7)^2 \left(1.66 - \frac{1}{2} \times 0.15\right)^2} = 27.295 \left[\frac{e^2}{\hbar}\right] \tag{15}$$

And the spin Hall conductivity $\sigma_s^e$ is about $\frac{\Delta_{so}}{E_g}$ of the valley Hall conductivity[5],

$$\sigma_s^e \approx \frac{0.15}{1.66} \sigma_v^e = 2.47 \left[\frac{e^2}{\hbar}\right] \tag{16}$$

We did not find an explicit expression like equation (14) of valley Hall conductivity $\sigma_v^e$ calculation in the literature at this time (although expression like equation(12) is given in like Ref. [5] ). However in Ref. [11] gives plot of Spin Hall conductivity as a function of $E_f$. The rough estimate in (16) is on order of magnitude agrees with FIG.2.

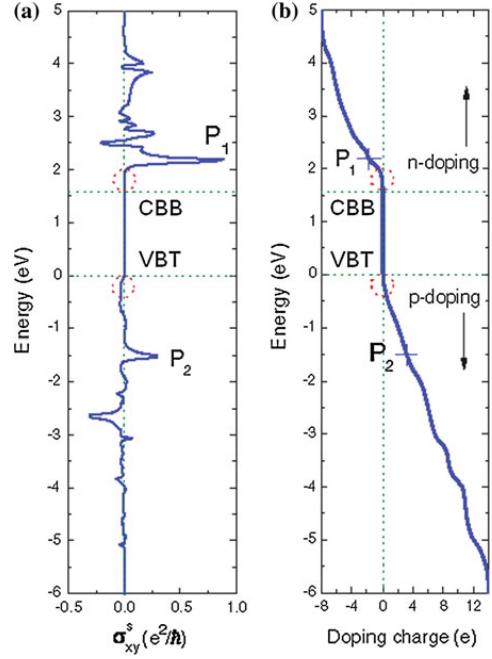

FIG.2. a The intrinsic spin Hall conductivity as a function of the Fermi energy for monolayer MoS2. b The n- and p-doping charge as a function of the Fermi energy.[9]

In summary, Barry curvature and its impact on Bloch electron transport was reviewed. For systems with either time-reversal or inversion symmetries broken, Barry curvature have important influence such as valley and spin Hall effect. After discussion of general symmetry and spin-orbital coupling Hamiltonian of monolayer $MoS_2$, an explicit expression for valley Hall conductivity $\sigma_v^e$ was derived with all available fitting parameters. With this expression $\sigma_v^e$ and $\sigma_s^e$ are calculated at $E_f = 2eV$ in the end as an example.

**References:**

[1] Griffiths, David (2004). *Introduction to Quantum Mechanics (2nd ed.)*. Prentice Hall. ISBN 0-13-111892-7

[2] D. Xiao, M.-C. Chang, and Q. Niu, *Rev. Mod. Phys. 82, 1959 (2010)*.
[3] A. H. Castro Neto, F. Guinea, N. M. R. Peres, K. S. Novoselov, and A. K. Geim, *Rev. Mod. Phys. 81, 109 (2009)*

[4] S. Lebègue and O. Eriksson, Phys. Rev. B 79, 115409 (2009)

[5] Di Xiao, Gui-Bin Liu, Wanxiang Feng, Xiaodong Xu, and Wang Yao, *Phys. Rev. Lett. 108, 196802 (2012)*

[6] Eugene Kadantsev, *Electronic Structure of Exfoliated MoS2*, Lecture Notes in Nanoscale Science and Technology Volume 21, 2014, pp 37-51

[7] L. F. Mattheiss, *Phys. Rev. B 8, 3719 (1973)*.

[8] Z.Y. Zhu, Y. C. Cheng, and U. Schwingenschlögl, *Phys. Rev. B 84, 153402 (2011)*

[9] Y. Liu, A. Goswami, F. Liu, D. L. Smith and P. P. Ruden, *J. Appl. Phys.* **116**, *234301 (2014)*

[10] A. Goswami, Y. Liu, F. Liu, P. P. Ruden, and D. L. Smith, *MRS Online Proceedings Library Archive 1505, (2013)*

[11] Feng, W.X., Yao, Y.G., Zhu, W.G., Zhou, J.J., Yao, W., Xiao, D, *Phys. Rev. B 86, 165108 (2012)*